\documentclass[doublecol]{epl2} 
\title{Stresses in smooth flows of dense granular media}
\shorttitle{Stresses in smooth granular flows }
\author{Martin Depken\inst{1} \and Jeremy B. Lechman\inst{2} \and Martin van Hecke\inst{3} \and Wim van Saarloos\inst{1} \and Gary S. Grest\inst{2}    }
\shortauthor{Martin Depken \etal}
\institute{
  \inst{1} Instituut--Lorentz,  Universiteit Leiden, Postbus
9506, 2300 RA Leiden, The Netherlands\\
  \inst{2} Sandia National Laboratories, Albuquerque, NM 87185,
USA\\
 \inst{3} Kamerlingh Onnes Lab, Universiteit Leiden, Postbus 9504,
2300 RA Leiden, The Netherlands
}


\abstract{The form of the stress tensor is investigated in smooth,
dense granular flows which are generated in split-bottom shear
geometries. We find that, within a fluctuation fluidized spatial
region, the form of the stress tensor is directly dictated by the
flow field: The stress and strain-rate tensors are co-linear. The
effective friction, defined as the ratio between shear and normal
stresses acting on a shearing plane, is found not to be constant
but to vary throughout the flowing zone. This variation can not be
explained by inertial effects, but appears to be set by the local
geometry of the flow field. This is in agreement with a recent
prediction, but in contrast with most models for slow grain flows,
and points to there being a subtle mechanism that selects the flow
profiles.
}

\begin{document}

\maketitle

\section{Introduction}
%MvH Rewrite
Granular media are amorphous and athermal materials which can jam
into stationary states, but which can also yield and flow
under sufficiently strong external forcing \cite{jamming,gm}.
Slowly flowing granulates, for which momentum transfer by enduring
contacts dominates over collisional transfer, are characterized by
a yielding criterion and rate independence. The former expresses
that granulates only start to flow when the applied shear stresses
exceed a critical yielding threshold~\cite{jamming,gm,nederman},
while the latter signifies that a change in the driving rate
leaves both the spatial structure of the flow and the stresses
essentially unaltered~\cite{bob,TCmueth,bonn,Fenistein,SFS}.

Solid friction exhibits a similar combination of yielding and
rate-independence: According to the Coulomb friction law, a block
of material resting on an inclined plane starts to slide when its
ratio of shear to normal forces exceeds the static friction
coefficient. And, once the block slides, the same ratio is given
by a lower dynamical friction coefficient, which is essentially
rate independent.

There is no unique manner in which these friction laws can be
translated into a continuum theory, and there exists a plethora of
approaches describing slow granular flows
\cite{nederman,SFS,tl,aranson,bazant,gdr,pouliquen_nature,Unger}.
To test these theories, one would like to determine the stresses
and strain rates within the material. However, experiments can not
easily access the flow in the bulk of the material, nor probe the
stress tensor in sufficient detail. In addition, slow grain flows
often exhibit sharp gradients, thus casting doubt on the validity
of continuum theories \cite{nederman,bob,TCmueth,bonn,tl}.
Finally, granular flows are notoriously sensitive to subtle
microscopic features \cite{TCmueth}, which often translates into a
substantial number of tunable parameters in the models
\cite{aranson}. As far as we are aware, no direct comparison
between the full stress and strain rate tensor has been undertaken
for slow granular flows.

\begin{figure}
\includegraphics[width=\columnwidth]{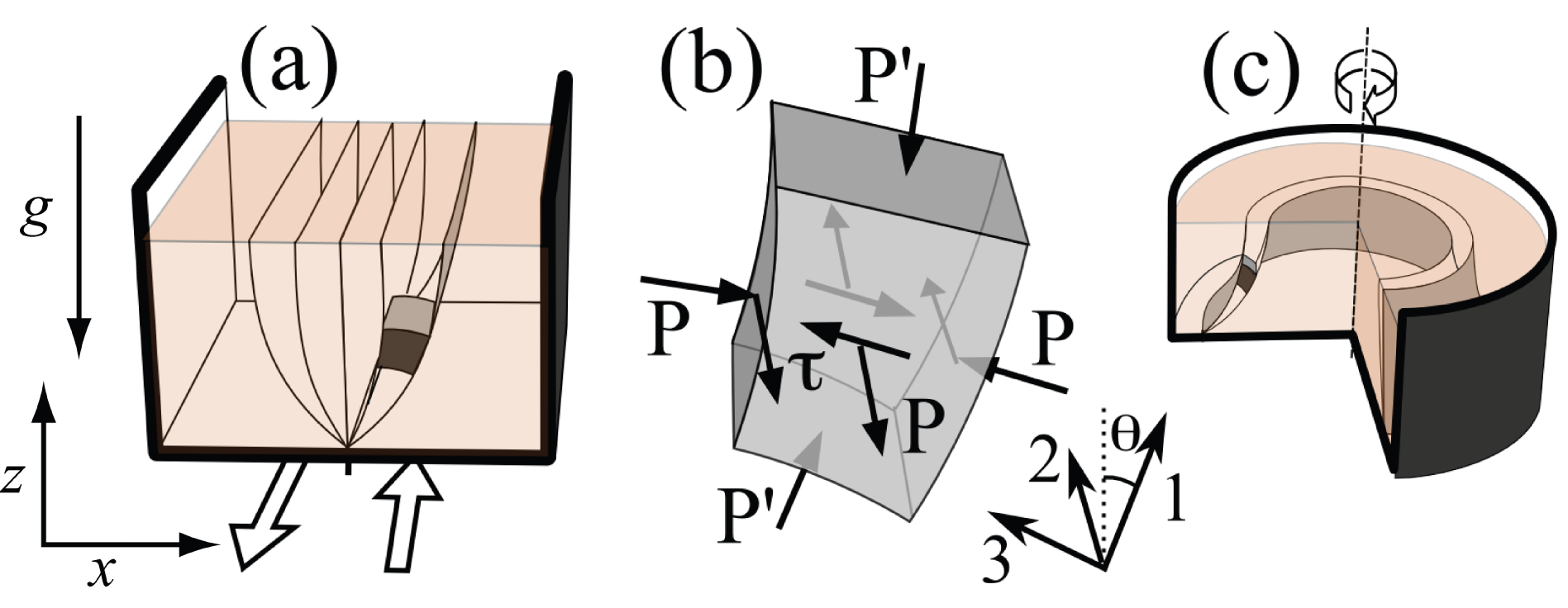}
\vspace{0cm} \caption{\label{fig:1}(a) Linear shear geometry where
a split along the middle of the system generates a wide shear zone
in a layer of grains. The curves indicate sheets of constant
velocity. (b) Cuboid element of material showing the definition of
the angle $\theta$, the stresses in the SFS framework, and the
labelling of the axis --- the grey objects in a) and c) are
examples of such elements. (c) Cylindrical split-bottom geometry,
where the grains are driven by the rotation of a bottom disc. The
two surfaces of rotation indicate sheets of constant angular
velocity. Note that in the limit $R_s \!\to\! \infty$ one obtains
the linear geometry (a).}
\end{figure}

In this Letter, we numerically study grain flows in split-bottom
geometries as shown in fig.~\ref{fig:1}. Recently, these flows
were shown to exhibit robust and continuum-like flow profiles that
are numerically tractable and are governed by a number of
universal, i.e. grain-independent, scaling relations, making them
eminently suitable for our purpose. We relate the stress tensor to
the strain-rate tensor in these flows, thus providing a benchmark
for the testing and development of theoretical models for smooth
and dense grain flows. Experiments and numerics so far have
focussed on the flow in a cylindrical geometry
(fig.~\ref{fig:1}c), where a wide shear zone is generated by
rotating a centre bottom disc with respect to the cylindrical
container \cite{Fenistein,Unger,xiang}. We present some data for
this cylindrical case, but focus on the linear version of this
geometry (fig.~\ref{fig:1}a), where we find a wide shear zone to
emanate from the relative motion of two bottom plates along their
``fault line''. In this system, the physics behind the stresses is
easier to disentangle because  the stream lines are not curved.

%\section{Main findings}
Our main finding is that, throughout the flowing zone, the stress
and strain tensor are co-linear, meaning that their
eigen-directions, or equivalently, their principle directions,
coincide. Moreover, we find that the ratios of the non-zero stress
components, such as the effective friction coefficient, which is
the ratio between shear and normal stresses acting on a shearing
plane, are not constants but vary throughout the flowing zone.
This variation is crucial to understand the finite width of the
shear zones, and is not due to the variation in the magnitude of
the local strain rate. Both of these findings are in accord with
the main features of theory developed in \cite{SFS}, and
constitute an important step forward in establishing a general
framework for the modelling of grains flows.

\section{SFS framework}
We formulate our results in the context of the theoretical
framework recently developed by Depken {\em et al.}~\cite{SFS}.
The central assumption of this so-called SFS theory is that, once
the material is flowing, strong fluctuations in the contact forces
enable otherwise jammed states to relax within a spatial region
which we refer to as the fluctuation fluidized region. In this
region there can not be a shear stress without a corresponding
shear flow. This assumption can be interpreted as stating that the
yielding threshold, which determines the onset of flow, is no
longer relevant once part of the material flows, since this
induces strong non-local fluctuations in the contact forces.
Further one observes that the flows can be locally (and in the
present cases also globally) seen as comprised of material sheets,
with no internal average strain rate, sliding past each other (see
fig.~\ref{fig:1}).

Combining these two ingredients, it follows that both the shear
strains and shear stresses in these material sheets are zero, and
we refer to them as a Shear Free Sheet (SFS). It also follows that
the stress and strain-rate tensors are co-linear. The major and
minor principle directions of the strain-rate tensor are at an
angle of 45$^\circ$ with respect to the SFSs, and in the more
intuitive basis specified by these sheets (see
fig.~\ref{fig:1}b) the stress tensor takes the form:
\begin{equation}
\sigma_{{\rm SFS}}=\left(\begin{array}{ccc}P'
&0&0\\0&P&\tau\\0&\tau&P\end{array}\right)~. \label{eq:stressr}
\end{equation}

To test this prediction, we check whether the numerically
obtained stresses are co-linear with the strain rate tensor and
thus are of the form (\ref{eq:stressr}). Moreover, when no further
assumptions are made, the three components $P$, $P'$, and $\tau$
will be different, and in general vary throughout the sample. In
fact, if the stress is of this form, a simple stress balance
argument shows that $\mu_{\rm eff}:=\tau/P$ has to vary throughout
the shear zones \cite{SFS}: A constant $\mu_{\rm eff}$ would
correspond to a shear zone of zero width, clearly inconsistent
with the available data \cite{Fenistein,xiang}.

To put these predictions in perspective, let us briefly consider
the case of faster flows, where collisions play a role. The
arguments for the form of the stress tensor can be extended to
apply also for such systems, and Pouliquen and co-workers
\cite{pouliquen_nature} have suggested that the stress is of the
form eq.~(\ref{eq:stressr}). However, they introduce the following
restriction: $P'\!=\!P$ and $\tau=\mu_{\rm eff}(I)P$, where the
effective friction is a material dependent function of the
so-called inertial number $I=\dot{\gamma} d /\sqrt{P/\rho}$
\cite{Inote}, and $d$ and $\rho$ are the particles diameter and
density, respectively~\cite{gdr,pouliquen_nature}. For the slow
flows under consideration here, we should consider the limit
$I\to0$. If we only consider $\mu_{\rm eff}$ to depend on $I$,
$\mu_{\rm eff}$ becomes a material constant, which is, as we
explained above, incompatible with the finite width of the shear
zones \cite{SFS,Unger,jop_private}. Our study will thus illuminate
how subtle details of the form of the stress tensor have
significant consequences for the grain flow.

\section{Method} The simulations are carried out with a discrete
element method (DEM) for $80\!-\!100$k mono-disperse Hertzian
spheres satisfying the Coulomb friction laws. The relevant
parameters describing the material properties of the spheres are
the normal stiffness $k_n=2 \times 10^5mg/d$, the tangential
stiffness $k_t=2/7k_n$, the normal and the tangential viscous
damping coefficients $\gamma _n=50 \sqrt{g/d}$ , $\gamma_t=0$, and
the microscopic coefficient of friction $\mu_m=0.5$. Here $d$ and
$m$ are the diameter and the mass of spheres, and $g$ is the
gravitational acceleration. The characteristic timescale $t_0$ is
given by $\sqrt{d/g}$ (e.g., $t_0 = 0.0101 sec$ if $d = 1 mm$). We
have studied a range of driving rates varying from from $\pm0.05$
to $\pm0.005~d/t_0$ and $0.015$ to $0.005~rad/t_0$ for the linear
and circular geometries, respectively. Stresses and velocities are
averaged over the symmetry direction (along split) and are
resolved with a resolution of $0.9d$ in the cross section. The
stress tensor within this volume is the sum of contact and
collisional stresses \cite{silbert_chute}, where the latter is
three orders of magnitude smaller than the former. The linear
setup has dimensions $20d$ in the shearing direction (periodic
boundary conditions), a width of $80d$, and a height of $50d$. The
details of the specific implementation can be found elsewhere
\cite{silbert_chute}.

\begin{figure}
\includegraphics[width=\columnwidth]{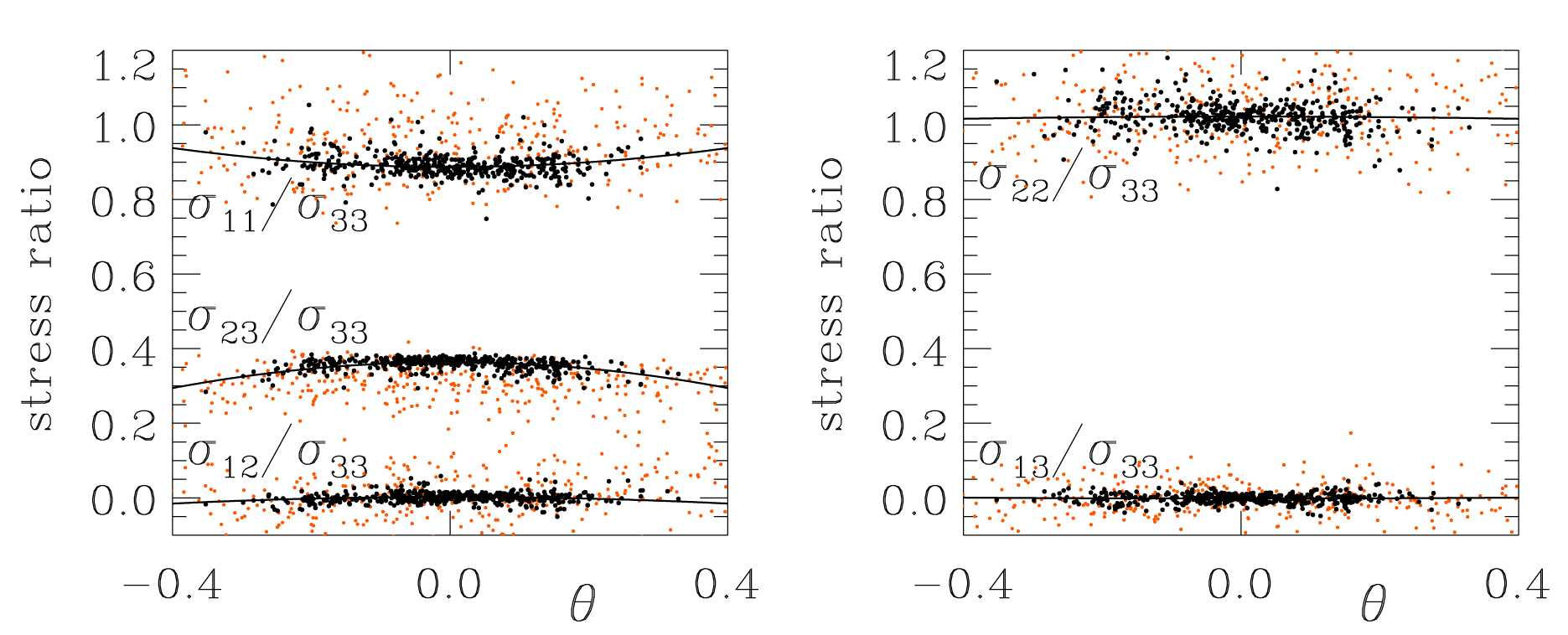}
\caption{\label{fig:2} Stress ratios $\sigma_{ij}/\sigma_{33}$ in
the linear geometry. The data was taken for a run where the
velocity difference across the sliding bottom plates was 0.05
$d/t_0$, and was averaged over the time interval ranging from 7150
$t_0$ to 9150 $t_0$ --- similar results where obtained for a
velocity difference that was ten times smaller. We plot the data
as a function of $\theta$, the angle between the ''1'' direction
of the SFS basis (see fig.~\ref{fig:1}) and the vertical. Black (red) dots
refer to points inside (outside) the fluctuation fluidized zone
(see text), and the curves are quadratic fits to the data in this
zone. The stress tensor follows eq.~(\ref{eq:stressr}): The ratios
$\sigma_{12}/\sigma_{33}$, $\sigma_{13}/\sigma_{33}$ and
$\sigma_{22}/\sigma_{33}$ are very close to zero, zero and one,
while the ratios $\sigma_{11}/\sigma_{33}=P'/P$ and
$\sigma_{23}/\sigma_{33}=\mu_{\rm eff}$ vary with $\theta$ and do
not attain any special value. }
\end{figure}

\section{The form of the stress tensor} We first study the
relation between stresses and strain rates in the linear geometry.
Through a cross section of the flow we record the time-averaged
stress and velocity fields, and from the latter we extract the
orientation of the SFS basis. In the region far away from the
shear zone, these fields fluctuate strongly, and we limit the
analysis to a ``fluctuation fluidized zone''. For this particular
data set we take the boundary of the fluctuation fluidized zone to
be defined by where the inertial number $I$ attains the value
$I_{\rm cut}=4\times10^{-5}$
--- why this is reasonable is detailed below.

Within this zone, we express the stresses in the SFS basis, and
compare our numerically obtained stresses to the SFS form
(\ref{eq:stressr}). We find that, due to gravity, all stress
components grow roughly proportional with depth. Since the SFS
theory makes no prediction on the absolute values of the various
stress components, we focus on the stress ratios
$\sigma_{ij}/\sigma_{33}$ (note that $\sigma_{23}/\sigma_{33}$
directly yields the effective friction coefficient, $\mu_{\rm
eff}$).

In fig.~\ref{fig:2} we plot the stress ratios as a function of the
angle $\theta$, which parameterizes the orientation of the SFS
basis with respect to gravity (see fig.~\ref{fig:1}b). Even though the
stress ratios could vary arbitrarily with position throughout the
cross section, we find that their main variation is with $\theta$
--- the relevance of this angle
will be discussed below. Figure~\ref{fig:2} illustrates that in
the fluctuation fluidized zone the stress tensor takes the SFS
form (\ref{eq:stressr}). First, all stress ratios within this
region appear to collapse on single curves when plotted as
function of $\theta$, while data outside the region is scattered
more strongly. Second, the values for the ratio's
$\sigma_{12}/\sigma_{33}$ and $\sigma_{13}/\sigma_{33}$ scatter
around zero. Third, the ratio $\sigma_{22}/\sigma_{33}$ is close
to one and does not vary with $\theta$. Together these points show
the validity of the SFS picture within the fluctuation fluidized
region (see below for a more precise definition). Finally, the
stress ratios $\sigma_{11}/\sigma_{33} = P'/P$ and
$\sigma_{23}/\sigma_{33} = \tau/P$ are not constant and do not
attain any special values. The data does not suggest that it is
possible to simplify the form of the stress tensor
(\ref{eq:stressr}) any further.

\section{Angle dependence of stress} The
variation of the effective friction $\mu_{\rm eff}$ with angle
$\theta$ takes on a special significance in the linear geometry.
In \cite{SFS} it was shown that, given a stress tensor of the SFS
form, force balance dictates that $\mu_{\rm eff}$ attains its
maximum in the middle of the shear zone, where $\theta\!=\!0$. It
was further shown that the curvature of $\mu_{\rm eff}(\theta)$
could be directly related to the scaling of the width of the shear
zone with vertical position in the sample; $W\sim z^\alpha$,
$\alpha=1/(1+\partial_{\theta\theta}\mu_{\rm eff}|_{\theta=0})$.
For constant $\mu_{\rm eff}$, $\alpha\!=\!1$ and  the shear zones
become of zero width \cite{SFS,Unger,jop_private}.

As fig.~\ref{fig:2}a shows, $\mu_{\rm eff}$ varies by roughly 10\%
throughout the fluctuation fluidized region and indeed attains a
maximum in the middle. A quadratic fit to $\mu_{\rm eff}$ yields
that $\partial_{\theta\theta}\mu_{\rm eff}|_{\theta=0} \approx
2.5[5]$, which suggests the scaling exponent $\alpha=0.35[5]$.
From the numerical data presented here, and from the data in
\cite{Fenistein} and \cite{xiang}, the value of this width
exponent can be estimated to be somewhere in the range $0.25-0.4$,
consistent with our estimate \cite{footnote_errorbar}. We
interpret this coincidence as a strong check on the validity of
the SFS form
--- the variation of $\mu_{\rm eff}$ is clearly a subtle effect, and one
could imagine that small and systematic deviations of the stresses
from the SFS form could destroy the relation between $\alpha$ and
$\partial_{\theta\theta}\mu_{\rm eff}|_{\theta=0}$.

\section{Spatial variation of the stress} 
\begin{figure*}[bht]
\includegraphics[width=\textwidth]{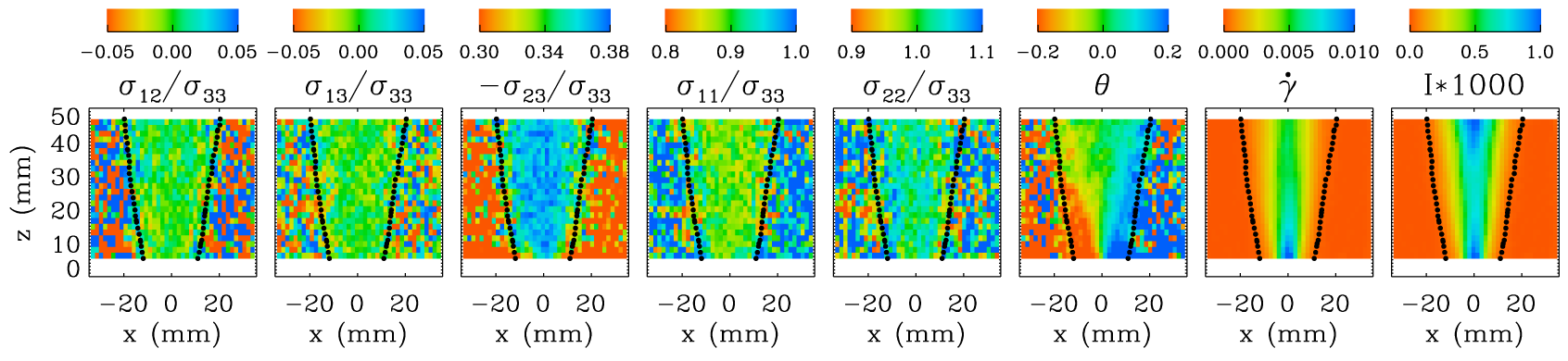}
\caption{Components of the stress tensor in the SFS basis for the
linear shearing geometry, compared to the angle $\theta$, local
strain rate $\dot{\gamma}$ and inertial number $I$. This data is
based on the same run as shown in fig.~\ref{fig:2}. A qualitative change of
behavior of the stress fields is clearly visible around the dashed
lines, which indicates the boundary of what is taken as the
fluctuation fluidized zone, $I=I_{\rm cut} = 4\times10^{-5}$ (see
text). \label{fig:3}}
\end{figure*}

In fig.~\ref{fig:3}
we plot the variations of the stress ratios $\sigma_{ij}/
\sigma_{33}$ throughout a cross section of the linear cell,
including data from outside the fluctuation fluidized zone. We
will now provide support for our assertion that the dominant
variations of the stress ratios are with $\theta$. We first
checked that the correlation between $\mu_{\rm eff}$ and
dimensionless quantities, such as the density and the curvatures
of the SFS basis, are unconvincing. Other potential candidates are
$\theta$, $\dot{\gamma}$, and $I$, and these are also shown
fig.~\ref{fig:3}. Figure~\ref{fig:3} shows that the spatial
variation of $\mu_{\rm eff}$ is closer to $\theta$ than it is to
$\dot{\gamma}$ or $I$.

Moreover, if the variation of $\mu_{\rm eff}$ was dominated by the
variation of $I$ or $\dot{\gamma}$, one would expect the width of
the shear zones to strongly depend on the shear rate
--- in stark contrast to both experimental \cite{Fenistein,xiang}
and numerical data \cite{xiang}. In fact, in runs done for a
driving rate which is a factor 10 smaller than shown here, the
stresses, flow profiles and $\mu_{\rm eff}(\theta)$ are
indistinguishable from those reported here --- the system is truly
rate independent. Finally, for the small inertial numbers here, $d
\mu_{\rm eff}/dI \sim {\cal O}(1)$ (based on the data presented
in~\cite{gdr}), while variations in $I$ over the shear zone are
${\cal O}(10^{-3})$ --- far too small to explain the 10\%
variation in $\mu_{\rm eff}$.

\section{The fluctuation fluidized region} Central to the SFS
picture is that shear flows induce force fluctuations that spread
through and fluidize the material, thus eliminating the yield
stress. How far do these fluidized zones spread? As can be seen in
Figs.~\ref{fig:2} and~\ref{fig:3}, there is a clear region within
which the stresses satisfy the SFS form of
eq.~(\ref{eq:stressr}), while outside this region the
stresses become noisy. We initially expected the total local
strain experienced since the start of the numerical experiment,
$\dot{\gamma} t$, to distinguish regions where fluctuations have
allowed the stresses to relax to the SFS form. But, when
attempting to maximize the spatial region of co-linearity, we
found that the inertial number, $I$, leads to a better estimate of
the fluctuation fluidized region: For the same required accuracy
in co-linearity, a larger region is selected~\cite{Inote}. In
Figs.~\ref{fig:2} and ~\ref{fig:3}, the region is defined as $I >
I_{\rm cut}= 4\times10^{-5}$. It should be noted that this cut-off
does not define a region of visible shear (as seen from the
$\dot{\gamma}$ plot in fig.~\ref{fig:3}), but rather a region
within which the microscopic fluctuations, created mainly in the
region of relatively large strain rates, remove any static shear
resistance.

Can we understand the numerical value of $I_{\rm cut}$? The total
strain experienced after $t=8000~ t_0$ (taken in the middle of the
total time-interval over which the stresses are averaged) equals
$8000 ~t_0~  \dot{\gamma} = 8000 ~\sqrt{d/g}~ \dot{\gamma}$. At
the edge of the fluctuation fluidized zone at a certain height
$z$, the local strain rate equals $I_{\rm cut}/d~
\sqrt{P/\rho}=I_{\rm cut}\sqrt{g(h-z)}/d$; hence the total strain
experienced at this edge equals $8000 ~I_{\rm
cut}~\sqrt{(h-z)/d}=0.32~ \sqrt{(h-z)/d}$. Near the bottom the
total strain thus approximates five, while near the surface it
becomes of order 0.3. Even though the fluctuation fluidized region
is not directly given by $\gamma$, these numbers nevertheless set
a reasonable scale for the amount of strain the material needs to
experience before it is fluidized, in particular if one realizes
that due to the pressure gradient, the strain near the bottom
couples more strongly to  the fluctuations of the forces near the
surface than vice versa. It should also be noted that we do not
expect the numerical value of $I_{\rm cut}$ to be universal --- in
particular, for longer runs we expect the fluctuation fluidized
region to spread slowly, with $I_{\rm cut} \sim 1/t $.

\section{Results in cylindrical geometry} 
\begin{figure}[tbh]
\begin{center}
\includegraphics[width=\columnwidth]{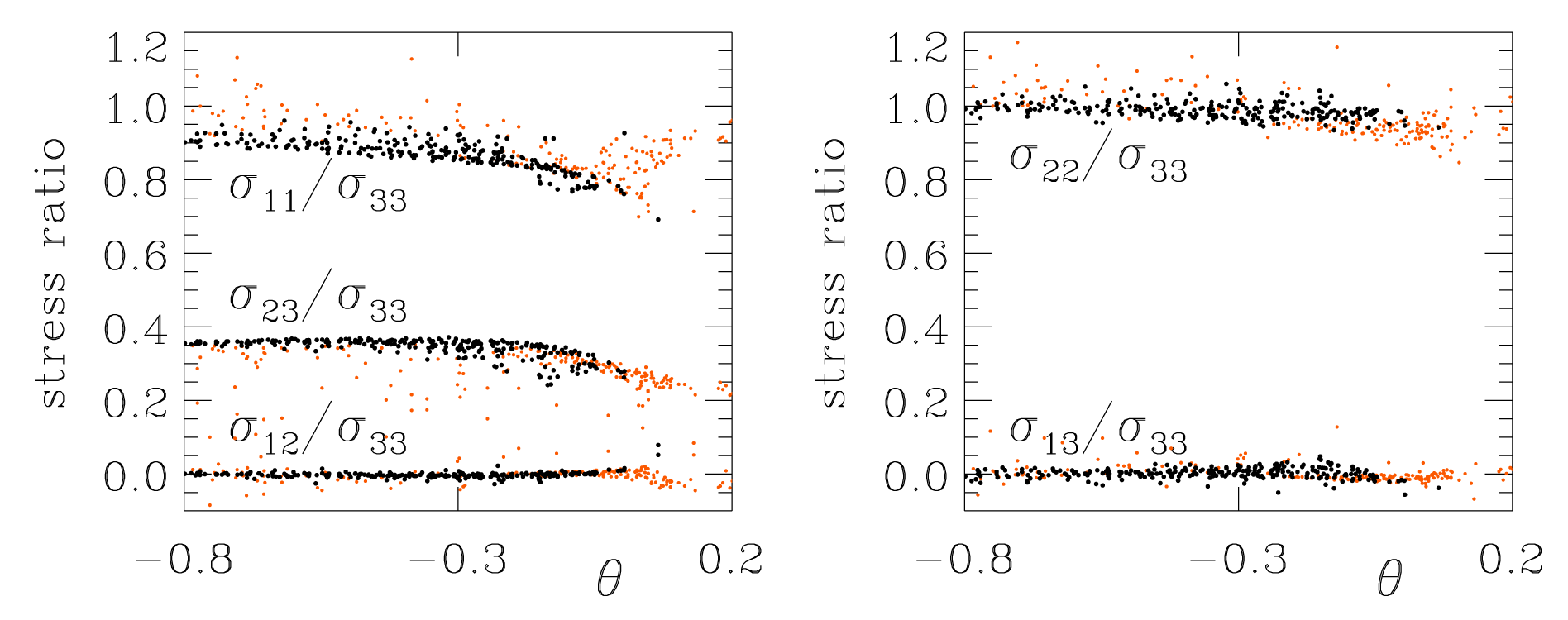}
\caption{ \label{fig:4} Stress ratios $\sigma_{ij}/\sigma_{33}$ as
a function of $\theta$, for a circular geometry. The driving rate
$\Omega$ was equal to 0.015 rad/$t_0$, and averages where taken
for time ranging from 8000 $t_0$ to $10^4$ $t_0$, corresponding to
the interval from approximately 19 to 24 turns. We have checked
that similar results where obtained for a run with $\Omega$ equal
to 0.005 rad/$t_0$. As before, black (red) dots referring to
points inside (outside) the fluctuation fluidized zone. For
details, see text.}
\end{center}
\end{figure}

In fig.~\ref{fig:4} we show simulation
results for a cylindrical geometry with $H/R_s = 0.675$ ---
similar results are reached for a number of other filling heights
not shown here \cite{the_future_is_bright}. Figure~4 shows that also
for the curved geometry, the stresses are in the SFS form: The
values for the ratio's $\sigma_{12}/\sigma_{33}$,
$\sigma_{13}/\sigma_{33}$ and $\sigma_{22}/\sigma_{33}$ scatter
around zero, zero, and one respectively, with the ratio's
$\sigma_{11}/\sigma_{33} = P'/P$ and $\sigma_{23}/\sigma_{33} =
\tau/P$ varying throughout the fluctuation fluidized zone.

Note that due to the more complex curved geometry, we have no
a-priori theoretical reason for expecting the stress ratios to
vary with $\theta$ alone. Moreover, there is no reason that
$\mu_{\rm eff}$ should be maximal in the middle, nor is it known
how $\mu_{\rm eff}(\theta)$ would be related to the width of the
shear zones  --- if at all.

\section{Conclusion and Outlook} Based on the single,
straightforward and minimal assumption that fluctuations on the
grain scale forbid the occurrence of shear stresses without an
associated shear flow, it was in \cite{SFS} predicted that the
stress tensor in slow grain flows should take the form
(\ref{eq:stressr}), with the stress ratios varying throughout the
sample. The data presented here fully confirms this prediction:
{\em{(i)}} In the flowing zones, the stresses indeed take the form
(\ref{eq:stressr}). The three different components $P'$, $P$ and
$\tau$ are sufficiently different that no further simplifications
are consistent with the data. {\em{(ii)}} The ratio $P'/P$ and the
effective friction $\mu_{\rm eff}=\tau/P$ are not constant.
{\em{(iii)}} The variation of $\mu_{\rm eff}$ can be directly
related to the width of the shear zones. {\em{(iv)}} For the
cylindrical geometry, the stress tensor also satisfies the SFS
criteria, with $P'/P$ and $\mu_{\rm eff}=\tau/P$ varying over the
shear zone, but due to the more complex geometry we can not relate
this variation directly to the width of the shear zones.

The SFS approach thus provides a powerful framework for unraveling
the relations between flow and stresses in granular media in
general, and the crucial but subtle spatial variation of the
effective friction $\mu_{\rm eff}$ and the unexpected variation of
$P'/P$ in particular.

The range of validity of the SFS approach is not yet clearly
mapped out, and additional studies to answer the following key
questions are called for. {\em{(i)}} How does the stress tensor
evolve when the flow rate is increased? The stress tensor in the
Pouliquen approach for fast flows is similar to ours, but with the
restrictions that $P'\!=\!P$ and that $\mu_{\rm eff}(I)$ is a
function of the local strain rate only \cite{pouliquen_nature}.
Here $\mu_{\rm eff}$ apparently depends on geometry, and the
crossover from geometry ($\theta$) to inertial number ($I$)
dependence needs to be explored.  {\em{(ii)}} We have seen here
that $P$ and $P'$ are systematically different, as was also seen
in simulations of chute flows \cite{silbert_chute}, and moreover,
that $P'/P$ is not a constant. Though we do not understand the
cause, nor the precise relevance, of this, it can not be a priori
ignored given the crucial role played by such variation of
$\mu_{\rm eff}$ in the formation of the wide shear zones in the
linear geometry. {\em{(iii)}} What distinguishes the zone where
the stresses are in the SFS form from the region where they are
not? Underlying the SFS picture is the assumption that the
fluctuations are sufficiently strong and fast, and one imagines
that far away from the shear zones this no longer holds true, thus
leading to a breakdown of co-linearity. Preliminary data suggest,
however, that the fluctuation fluidized region, most of which is
established after a short transient, very slowly expands as a
function of time \cite{the_future_is_bright}. Possibly, after
sufficiently long time, all the material has experienced flow and
the stress tensor takes the SFS form everywhere, but this may be
hard to verify numerically. Similar questions on the validity of
the SFS framework can also be asked when the driving rate is made
excessively slow. Ultimately, these questions are related to the
puzzling nature of the transition between the static and flowing
state of granular media \cite{silbert_chute,corwin}. {\em{(iv)}}
Is the variation of the effective friction the cause or effect of
the smoothness of our shear profiles? We suggest that the
spreading of contact force {\em fluctuations}, from the rapidly
shearing center to the tails of the shear zones, may elucidate the
microscopic mechanism by which the width of the shear zones are
selected. In this picture, the spread of fluctuations would also
drive the subtle variations of the coarse grained and time
averaged stresses, which thus serve to signal an underlying, but
unknown, fluctuation driven mechanism \cite{unger2006}.

\acknowledgments
We thank M. Cates for discussions. MD
acknowledges support from the physics foundation FOM and the EU
network PHYNECS, and MvH from the science foundation NWO through a
VIDI grant. Sandia is a multiprogram laboratory operated by Sandia
Corporation, a Lockheed Martin Company, for the United States
Department of Energy's National Nuclear Security Administration
under contract DE-AC04-
94AL85000.


\begin{thebibliography}{0}
\bibitem{Inote} The inertial number can be seen as the ratio of
the characteristic timescale for a grain to relax back from a
local dilation event ($d\sqrt{\rho/P}$) to the strainrate
($1/\dot\gamma$).
\bibitem{footnote_errorbar}
The error bar on $\partial_{\theta\theta}\mu_{\rm eff}|_{0}$ is
based on its variation with $I_{\rm cut}$, when this cutoff ranges
from $1.5 \times 10^{-5}$ to $10^{-4}$.
\end{thebibliography}
\end{document}